\documentclass[twocolumn,english,aps.two coloumn]{revtex4-1}
\usepackage{ae,aecompl}
\usepackage{beramono}
\usepackage[T1]{fontenc}
\usepackage[latin9]{inputenc}
\usepackage{amssymb}
\usepackage{graphicx}

\makeatletter

\providecommand{\tabularnewline}{\\}

\makeatother

\usepackage{babel}
\begin{document}

\title{Universal conductance fluctuations and direct observation of crossover
of symmetry classes in topological insulators}

\author{Saurav Islam$^{1}$, Semonti Bhattacharyya$^{1,2}$, Hariharan Nhalil$^{1}$,
Suja Elizabeth$^{1}$, Arindam Ghosh$^{1,3}$}

\affiliation{$^{1}$Department of Physics, Indian Institute of Science, Bangalore: $560012$.}

\address{$^{2}$School of Physics and Astronomy, Monash University, VIC $3800$,
Australia.}

\affiliation{$^{3}$Center for Nanoscience and Engineering, Indian Institute of
Science, Bangalore: $560012$.}
\email{isaurav@iisc.ac.in}

\begin{abstract}
A key feature of disordered topological insulators\ (TI) is symplectic
symmetry of the Hamiltonian which changes to unitary when time reversal
symmetry is lifted and the topological phase transition occurs. However,
such a crossover has never been explicitly observed, by directly probing
the symmetry class of the Hamiltonian. In this report, we have probed
the symmetry class of topological insulators by measuring the mesoscopic
conductance fluctuations in the TI Bi$_{1.6}$Sb$_{0.4}$Te$_{2}$Se,
which shows an exact factor of two reduction on application of a magnetic
field due to crossover from symplectic to unitary symmetry classes.
The reduction provides an unambiguous proof that the fluctuations
arise from the universal conductance fluctuations\ (UCF), due to
quantum interference and persists from $T\sim22$\ mK to $4.2\ $K.
We have also compared the phase breaking length $l_{\phi}$ extracted
from both magneto-conductivity and UCF which agree well within a factor
of two in the entire temperature and gate voltage range. Our experiment
confirms UCF as the major source of fluctuations in mesoscopic disordered
topological insulators, and the intrinsic preservation of time reversal
symmetry in these systems. 
\end{abstract}
\maketitle
Topological insulators\ \cite{hasan2010colloquium,moore2010birth,bernevig2006quantum,konig2007quantum}
at zero magnetic field are time reversal invariant systems characterized
by surface states with a linear band structure. The Hamiltonian for
such surface states is described by $H=\hbar v_{F}\overrightarrow{\sigma}\cdot\overrightarrow{k}$
which belongs to the AII/symplectic universality class, where $v_{F}$,
$\vec{\sigma}$, and $\vec{k}$ are the Fermi velocity, spin matrices,
and momentum respectively. This is also known as the Anderson universality
class for non-relativistic particles in the presence of a random spin-orbit
coupling where time reversal symmetry\ (TRS) is preserved\ \cite{adroguer2012diffusion}.
The addition of an external magnetic field or ferromagnetic impurities
introduce a Zeeman/orbital term in the Hamiltonian and breaks the
TRS which results in a topological to trivial phase transition in
the bulk states and manifests as a gap opening in the linear surface
states\ \cite{bernevig2006quantum}. In terms of random matrix theory,
this crossover at the surface states is well described by a crossover
from AII/symplectic to A/unitary class in the system. Experimentally,
the sensitivity of transport to magnetic impurities\ \cite{chang2013experimental,bao2013quantum}
or the saturation of the phase breaking length at low temperatures
are directly connected to the TRS in TI systems\ \cite{liao2017enhanced,bao2012weak}.
This makes an explicit demonstration of the symplectic to unitary
crossover an important task, which has however, not been achieved
yet. 

One direct method to probe such crossover of symmetry classes is universal
conductance fluctuations (UCF)\ \cite{lee1985universal,feng1986sensitivity,altshuler1986repulsion},
which is observed in mesoscopic devices, when the length of the sample
$L$ becomes comparable to $l_{\phi}$, the phase breaking length.
UCF is an effect which results from quantum interference of all possible
electron paths traversed between two points in a sample making the
electrical conduction sensitive to the Fermi energy, magnetic field
and impurity configuration. These fluctuations are independent of
the specific material properties or geometry, and is determined by
the physical symmetries of the Hamiltonian. Within the framework of
random matrix theory, the magnitude of UCF is proportional to\ \cite{altshuler1986repulsion}

\begin{equation}
\langle\delta G^{2}\rangle\propto\left(\frac{e^{2}}{h}\right)^{2}\frac{ks^{2}}{\beta}\label{eq:UCF RMT}
\end{equation}
Here $\beta$, $s$, and $k$ are the Wigner-Dyson parameter, Kramer's
degeneracy, and the number of independent eigen modes of the Hamiltonian
respectively\ (Table\ \ref{Table 1}). This UCF based technique
has been previously used as an experimental probe in mesoscopic samples
of graphene\ \cite{geim2004,neto2009electronic}, where a factor
of four reduction was observed in the magnitude as a function of number
density due to crossover from symplectic to orthogonal classes\ \cite{pal2012direct}.
Similar reduction of noise with magnetic field were observed in metal
films\ \cite{mcconville1993weak,birge1989electron,birge1990conductance,moon1996observation},
metallic single crystals of silicon\ \cite{ghosh2000universal},
and also in $\delta$-doped silicon-phosphorous systems\ \cite{shamim2017dephasing,shamim2014spontaneous}.
Though a symmetry class crossover in TI has been induced by addition
of ferromagnetic impurities and inferred from weak anti-localization\ \cite{he2011impurity},
a more direct observation of the symmetry class and its crossover
on breaking TRS, which does not require any addition of impurity remains
experimentally elusive. In this report, we present results of conductance
fluctuation measurements in topological insulator Bi$_{1.6}$Sb$_{0.4}$Te$_{2}$Se
on atomically thin hexagonal-boron nitride\ (hBN) substrate which
shows a factor of two reduction on application of a magnetic field,
thereby suggesting that the fluctuations indeed arise from UCF and
the reduction is driven by a crossover from symplectic to unitary
symmetry class. We have also extracted the phase breaking length from
both magneto-conductance\ ($l_{\phi}^{MR}$) and universal conductance
fluctuations\ ($l_{\phi}^{UCF}$) and found a close agreement.

\begin{figure*}
\includegraphics{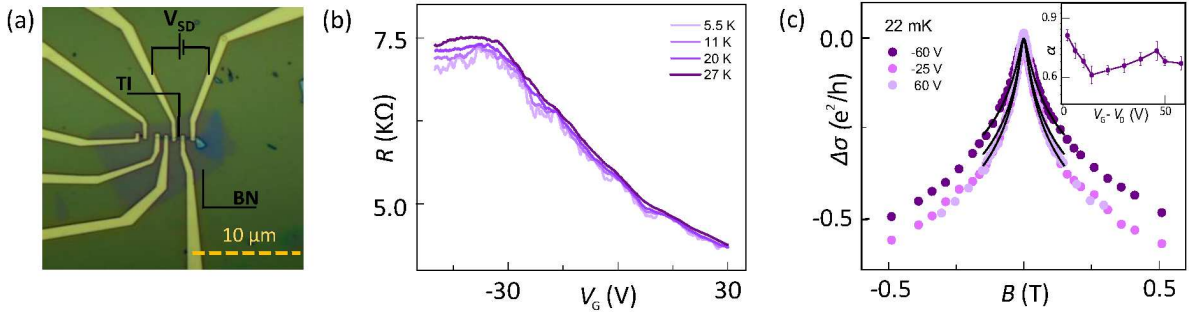}
\label{Fig1}
\caption{\textbf{Basic characteristics of TI field effect transistor\ (FET).}\ (a)\ Optical
micrograph of a typical TI on BN FET device. (b)\ $R$-$V_{G}$ of
the device at different temperatures. (c)\ MR at three different
gate voltages at $T=22$\ mK. The solid lines are fits to the data
according to the Eq.\ \ref{eq:HLN} (inset shows $\alpha$ extracted
from the fits using Eq.\ \ref{eq:HLN}).}
\end{figure*}

\begin{figure}
\includegraphics{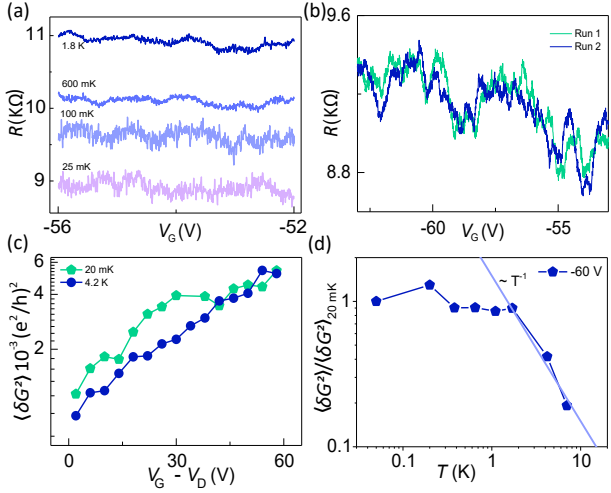}
\label{Fig2}
\caption{\textbf{Features of conductance fluctuations}.\ (a)\ $R$-$V_{G}$
in a small gate voltage window of $4$\ V for four different temperatures
used to calculate $\langle\delta G^{2}\rangle$ by using a smooth
polynomial fit. The curves have been offset for clarity. (b)\ $R$-$V_{G}$
for two different runs showing the reproducibility of conductance
fluctuations arising due to quantum interference. (c)\ Magnitude
of conductance fluctuations as a function of gate voltage showing
a monotonic increase away from the Dirac point. (d)\ $T$-dependence
of $\langle\delta G^{2}\rangle$ normalized by the magnitude at $20$\ mK
showing a gradual decrease with increasing $T$, a characteristic
feature of UCF. The solid line shows $\langle\delta G^{2}\rangle\sim1/T$
for $T>2$\ K. }
\end{figure}

The devices studied in this paper were fabricated from a $11$\ nm
thick topological insulator Bi$_{1.6}$Sb$_{0.4}$Te$_{2}$Se\ \cite{taskin2011observation}
exfoliated on SiO$_{2}$/Si wafer and then transferred onto a $14$\ nm
hBN substrate. The heterostructure was then finally transferred onto
a heavily doped SiO$_{2}$/Si substrate with the $285$\ nm thick
SiO$_{2}$ acting as a back gate dielectric, using a home-made transfer
technique. hBN was used to reduce the effect of dangling bonds and
charged traps of the SiO$_{2}$ substrate on the electrical transport
in the TI channel\ \cite{dean2010boron,karnatak2016current}. The
quarternary alloy Bi$_{1.6}$Sb$_{.4}$Te$_{2}$Se offers reduced
bulk number density due to compensation doping resulting in enhanced
surface transport\ \cite{taskin2011observation}. The contact pads
were defined by standard electron-beam lithography followed by metallization
using $5/40$\ nm Cr/Au (Fig.\ 1(a)). The sample was coated with
a layer of PMMA\ (poly(methylmethacrylate)) during the entire measurement
cycle. All measurements from $22$\ mK to 4.2\ K were done in a
dilution refrigerator. Resistivity measurements were performed using
a low frequency AC-four probe technique with carrier frequency of
$18$\ Hz. The excitation current was $0.1$\ nA for most of the
measurements to reduce the effect of Joule heating except at $4.2$\ K
when it was increased to $1$\ nA. The resistance\ ($R$) vs gate
voltage\ ($V_{G}$) is shown in Fig.\ 1(b), where a maximum in the
resistance at $V_{G}\approx-38$\ V at $5.5$\ K represents the
Dirac point. The number density calculated at $V_{G}=0$\ V using
$n=\frac{C_{S}\left(V_{G}-V_{D}\right)}{e}$ is $-2.9\times10^{16}$\ m$^{-2}$.
Here $C_{S}$ is the series capacitance of SiO$_{2}$ and hBN layers.
Fig.\ 1(c) shows weak-antilocalization phenomenon characterized by
a cusp in the quantum correction to conductivity $\triangle\sigma$
at $B=0$\ T. Spin momentum locking in TI leads to an additional
$\pi$ Berry phase between the backscattered, time reversed path of
the carriers, leading to negative magneto-conductance, a signature
of the symplectic phase. The magneto-conductance data can be fitted
with the Hikami-Larkin-Nagaoka (HLN) equation for diffusive metals
with high spin orbit coupling\ $(\tau_{\phi}>>\tau_{so},\tau_{e})$\ \cite{hikami1980spin,bao2012weak}:

\begin{table}
\caption{Values of symmetry parameters for the two classes relevant for TI.}
\label{Table 1}%
\begin{tabular}{|c|c|c|c|c|c|c|}
\hline 
Ensemble & TRS & $k$ & $s$ & $\beta$ & $H_{ij}$ & $\langle\delta G^{2}\rangle\propto\left(\frac{e^{2}}{h}\right)^{2}\frac{ks^{2}}{\beta}$\tabularnewline
\hline 
\hline 
Symplectic & Yes & $1$ & $2$ & $4$ & real quarternion & $1$\tabularnewline
\hline 
Unitary & No  & $1$ & $1$ & $2$ & complex & $0.5$\tabularnewline
\hline 
\end{tabular}
\end{table}

\begin{figure*}
\includegraphics{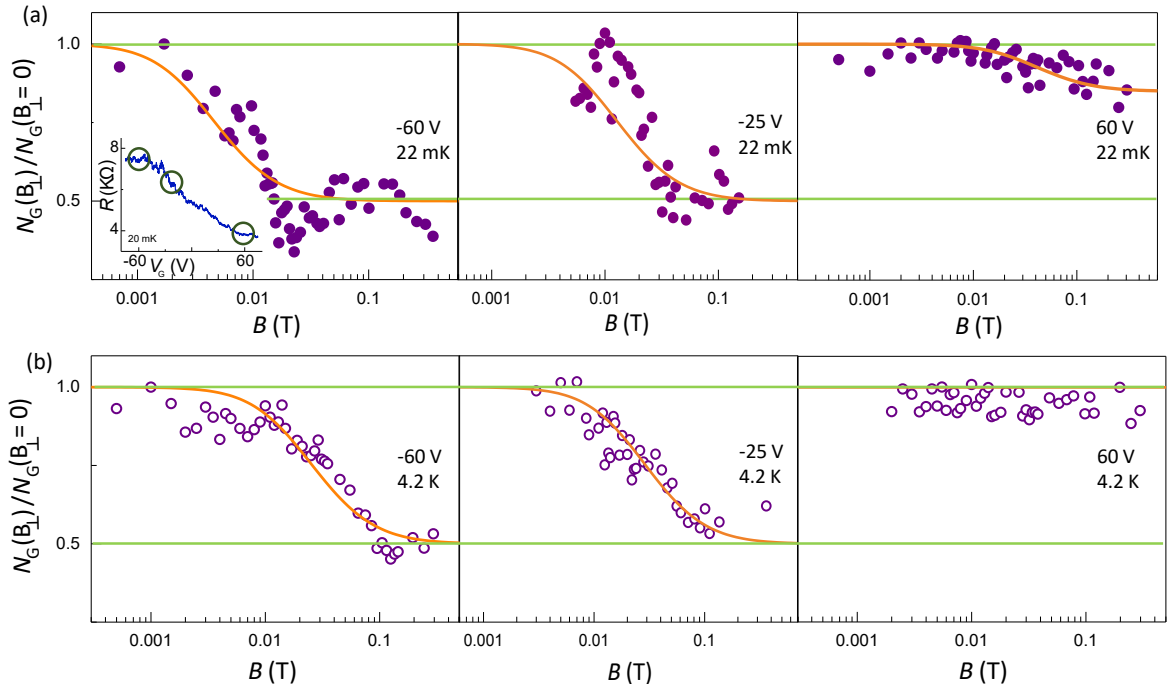}
\label{Fig3}
\caption{\textbf{Magnetic field dependence of normalized UCF magnitude.}\ (a)\ $v(B_{\perp},T)=N_{G}(B_{\perp})/N_{G}(B_{\perp}=0)$
for three different gate voltages\ ($-60$\ V, $-25$\ V and $60$\ V)
for $22$\ mK clearly exhibiting a factor of two reduction which
diminishes as the Fermi energy is tuned towards the bulk. The solid
lines indicate fit to the data using Eq.\ \ref{eq: UCF Fitting}.
The noise data for $60$\ V at $T=20$mK has been fitted using the
crossover function $\nu^{\prime}(B,T)$\ (See supplementary). Inset
in (a)\ shows $R$ vs $V_{G}$ at $T=20$\ mK. The circled regions
indicate windows where UCF was measured. (b)\ Normalized UCF magnitude
three gate voltages$-60$\ V, $-25$\ V and $60$\ V\  at $4.2$\ K.
The solid lines indicate fit to the data using Eq.\ \ref{eq: UCF Fitting}.}
\end{figure*}

\begin{equation}
\triangle\sigma=-\alpha\frac{e^{2}}{\pi h}\left[\psi\left(\frac{1}{2}+\frac{B_{\phi}}{B}\right)-\ln\left(\frac{B_{\phi}}{B}\right)\right]\label{eq:HLN}
\end{equation}

where $\tau_{\phi}$, $\tau_{so}$, $\tau_{e}$ are the phase coherence
or dephasing time, spin-orbit scattering time and elastic scattering
time respectively, $\psi$ is the digamma function and $B_{\phi}$
is the phase breaking field. Here $\alpha$ and $B_{\phi}$ are the
fitting parameters. The phase coherence length\ ($l_{\phi}^{MR}$)
can be extracted using $l_{\phi}^{MR}=\sqrt{\hbar/4eB_{\phi}}$. The
value of $\alpha$ gives an estimate of the number of independent
conducting channels in the sample. $\alpha=0.5$ indicates a single
transport channel whereas a value of $1$ indicates two independent
channels contributing to magneto-transport. In our case, the value
of $\alpha$ at $T=22$\ mK (inset of Fig.\ 1(c)) varies from $0.8$
near the Dirac point ($V_{G}=-60$\ V) and gradually reduces to a
value of around $0.6$ for more positive gate voltages. This probably
indicates that as the Fermi energy is tuned away from the charge neutrality
point, the bulk carriers start contributing to transport resulting
in a reduction in the value of $\alpha$, signifying coupling of bulk
and surface transport\ \cite{liao2017enhanced}.

The magnitude of conductance fluctuations\ $\langle\delta G^{2}\rangle$,
is evaluated using a method similar to Ref\ \cite{gorbachev2007weak,pal2012direct}
by varying the Fermi energy with the back gate voltage in steps of
$5$\ mV over a small window of $4$\ V so that statistically meaningful
data\ (about\ $800$ realizations)\ are recorded, without changing
the conductance appreciably in a two terminal configuration for each
transverse magnetic field. $\langle\delta G^{2}\rangle$ is extracted
from $R$-$V_{G}$ by fitting the data with a smooth polynomial curve\ \cite{pal2012direct,gorbachev2007weak}.
The variance of the residual of the fit gives the value of $\langle\delta R^{2}\rangle$
and the mean value corresponds to $\langle R\rangle$. $\langle\delta G^{2}\rangle$
is then obtained using the relation: $\langle\delta G^{2}\rangle=\langle\delta R^{2}\rangle/\langle R\rangle^{4}$.

Fig.\ 2(a) shows typical $R$-$V_{G}$ sweeps at four different temperatures
for the device, where the fluctuations decrease with $T$, a hallmark
of UCF. The run to run reproducibilty as a function of $V_{G}$ in
Fig.\ 2(b) further confirms the aperiodic yet reproducible nature
of the fluctuations. The $V_{G}$-dependence of $\langle\delta G^{2}\rangle$
is shown in Fig\ 2(c). The magnitude of $\langle\delta G^{2}\rangle$
shows an increase as the Fermi energy is tuned away from the electron-hole
puddle dominated regime towards higher number densities. Such behavior
is a unique signature of Dirac Fermionic systems like TI surface states
where the disorder potential due to charged impurities is long range
in nature\ \cite{pal2012direct,rossi2012universal}. At more positive
gate voltages, the value of $\langle\delta G^{2}\rangle$ at $T=20$\ mK
and $4.2$\ K are similar. This may be due to increased contribution
of scattering due to bulk defects which are the dominant source of
noise in TI\ \cite{bhattacharyya2015bulk,bhattacharyya2016resistance,islam2017bulk,pelz1987quantitative,hershfield1988sensitivity}.
The temperature dependence of conductance fluctuations (Fig.\ 2(d)),
shows an increase as $T$ is reduced to $2$\ K, below which it saturates.
The saturation of $\langle\delta G\rangle^{2}$ at $T<2$\ K can
be due to saturation of $l_{\phi}$. Such saturation has been previously
seen in various systems and can arise due to the presence of magnetic
impurities in the system\ \cite{schopfer2003anomalous} or when the
spin orbit length becomes comparable to the phase breaking length\ \cite{fukai1990saturation}.
For $T>2$\ K, the magnitude of $\langle\delta G^{2}\rangle$ decreases
with increasing $T$ as $\langle\delta G^{2}\rangle\propto1/T$ which
can be explained from the dependence of $\langle\delta G^{2}\rangle$
on $l_{\phi}$ and the number of active fluctuators\ ($n_{s}$).
For $T\rightarrow0$, UCF magnitude $\left\langle \left(\delta G\right)^{2}\right\rangle ^{\frac{1}{2}}\rightarrow e^{2}/h$,
while at finite temperature\ \cite{birge1990conductance,feng1986sensitivity,shamim2017dephasing},

\begin{equation}
\langle\delta G^{2}\rangle\simeq\left(\frac{e^{2}}{h}\right)^{2}\alpha(k_{F}\delta r)\frac{1}{k_{F}l}\frac{L_{y}}{L_{x}^{3}}n_{s}(T)l_{\phi}^{4}\label{eq:UCF Mag}
\end{equation}
where $k_{F}$, $l$, $L_{x}$ and $L_{y}$ are the Fermi wave-vector,
mean free path and sample dimensions in $x$ and $y$ directions respectively.
$\alpha(x)$ represents the change of the phase of electron wave-function
due to scattering by a moving impurity at a distance $\delta r$.
Assuming electron-electron interaction mediated dephasing, $l_{\phi}^{2}\propto1/T$
and $n_{s}(T)\propto T$\ \cite{birge1990conductance,feng1986sensitivity,shamim2017dephasing,altshuler1982effects},
we have $\langle\delta G^{2}\rangle\propto l_{\phi}^{4}n_{s}(T)\propto1/T$\ (Fig.\ 2(d)),
as observed at $T>2$\ K.

\begin{figure}
\includegraphics[width=1\columnwidth]{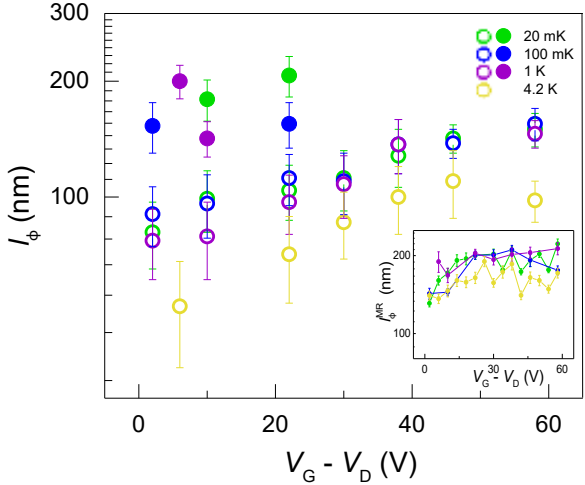}
\label{Fig 4}
\caption{\textbf{Comparison of phase breaking length\ ($l_{\phi}$) from different
methods.}\ $l_{\phi}$ extracted from UCF as a function of $V_{G}$
for different $T$ from magnetic field dependence using Eq.\ \ref{eq: UCF Fitting}\ (solid
circles) and directly at $B_{\perp}=0$\ T using Eq.\ \ref{eq:ucf 2}\ (hollow
circles). Inset shows $l_{\phi}^{MR}$ extracted from magneto-conductance
using Eq.\ \ref{eq:HLN} as a function of $V_{G}$ for different
$T$. $l_{\phi}$ obtained from three different methods agree well
within a factor of two.}
\end{figure}

 In order to (a)\ conclusively establish the role of UCF, and (b)\ investigate
the crossover in the symmetry class directly, we have measured $\langle\delta G^{2}\rangle$
as a function of perpendicular magnetic field\ ($B_{\perp}$) at
fixed $V_{G}$. The magnitude of the conductance fluctuations is plotted
as $v(B_{\perp},T)=N_{G}(B_{\perp})/N_{G}(B_{\perp}=0)$ where $N_{G}=\langle\delta G^{2}\rangle/\langle G^{2}\rangle$
is the normalized variance. As a function of increasing $B_{\perp}$,
as shown in Fig.\ 3(a) and Fig.\ 3(b) at $T=22$\ mK and $T=4.2$\ K,
we observe a clear factor of two reduction in the UCF magnitude at
$-60$\ V and $-25$\ V whereas at $60$\ V, the reduction significantly
reduced. For gate voltages closer to the Dirac Point\ ($-60$\ V
and $-25$\ V), the reduction occurs for $B_{\perp}\sim0.01-0.1$\ T,
which is similar to the field scales for quantum interference effect\ ($B_{\phi}$)
reported for TI\ \cite{liao2017enhanced,zhang2012magneto}. Since
$B_{\phi}\sim\hbar/4el_{\phi}^{2}$ , the increase in field scales
with $T$ can be readily attributed to the decrease in $l_{\phi}$
with increasing $T$. Although $l_{\phi}\propto T^{-1/2}$ expected
from electron-electron scattering would lead to much larger change
in $B_{\phi}$ in the experimental temperature range, we believe a
saturation in $l_{\phi}$\ \cite{schopfer2003anomalous,fukai1990saturation}
limits the decrease in $B_{\phi}$ at low temperatures. The absence
of a clear reduction by a factor of two at high positive voltages
can be due to additional noise contributions in the system. As $E_{F}$
is tuned towards the bulk bands, trapping-detrapping processes from
the charged impurities in the bulk, which are independent of $B_{\perp}$,
and are known to be dominant source of noise in TI increases\ \cite{bhattacharyya2015bulk,bhattacharyya2016resistance,islam2017bulk,pelz1987quantitative,hershfield1988sensitivity},
which diminishes the factor of two reduction. 

For a quantitative understanding, we have fitted the normalized magnitude
with the expression\ \cite{stone1989reduction,shamim2017dephasing,lee1987universal}

\[
v(B,T)=\frac{1}{2}+\frac{1}{b^{2}}\sum_{n=0}^{\infty}\frac{1}{[\left(n+\frac{1}{2}\right)+\frac{1}{b}]^{3}}
\]
\begin{equation}
=\frac{1}{2}-\Psi^{''}\frac{1}{2b^{2}}\left(\frac{1}{2}+\frac{1}{b}\right)\label{eq: UCF Fitting}
\end{equation}
Here $b=8\pi B(l_{\phi})^{2}/(h/e)$ and $\Psi^{''}$ is the double
derivative of the digamma function and $l_{\phi}$ is the fitting
parameter. The solid lines in Fig.\ 3 are fits according to Eq.\ \ref{eq: UCF Fitting},
which capture the variation of noise magnitude with $B_{\perp}$ well,
especially at gate voltage values close to the Dirac point. The corresponding
$l_{\phi}$, extracted from fitted value of $b$, are plotted in Fig.\ 4\ (solid
circles). At $V_{G}=60$\ V, the overall reduction allows us to estimate
$\thickapprox28$\ \% of the observed noise magnitude to arise from
UCF at $T=20$\ mK\ (Fig.\ 3(a)) but becomes negligibly small at
higher $T$\ (Fig.\ 3(b)).

Finally, we have evaluated the $V_{G}$-dependence of $l_{\phi}$
from three different methods (a)\ $\langle\delta G^{2}\rangle$ magnitude,
(b)\ $B_{\perp}$-dependence of noise, and (c)\ magneto-conductance.
We have however, restricted the calculation between $V_{G}=-60$\ V
to $0$\ V, where UCF is the dominant source of noise. $l_{\phi}$
extracted directly from $\langle\delta G^{2}\rangle$ at $B_{\perp}=0$\ T
using the expression\ \cite{adroguer2012diffusion,beenakker1991quantum,akkermans2007mesoscopic}:
\begin{equation}
\langle\delta G^{2}\rangle\simeq\frac{3}{\pi}\left(\frac{e^{2}}{h}\right)^{2}\left(\frac{l_{\phi}}{L}\right)^{2}\label{eq:ucf 2}
\end{equation}
is shown in Fig.\ 4\ ($L=W$\ (width)$=2$\ $\mu$m)\ (hollow
circles), while $l_{\phi}^{MR}$ extracted from magneto-conductance
using Eq.\ \ref{eq:HLN}, is shown in the inset of Fig.\ 4. We find
that $l_{\phi}^{MR}$ and $l_{\phi}^{UCF}$\ (obtained from both
Eq.\ \ref{eq: UCF Fitting} and Eq.\ \ref{eq:ucf 2}) are similar
in magnitude within a factor of two, rendering validity to the factor
of two reduction and the corresponding analysis. $l_{\phi}^{MR}$
increases away from the Dirac point, similar to the trend of $l_{\phi}$
obtained from UCF\ (Fig.\ 4). Near the Dirac point, the system is
highly inhomogeneous with the presence of electron-hole puddles. As
the gate voltages is tuned away from the charge neutrality point,
the carrier concentration increases leading to enhanced screening
which suppresses dephasing due to electromagnetic fluctuations. This
leads to a larger phase breaking length away from the Dirac point\ \cite{tian2014quantum,chiu2013weak,martin2008observation}.
The factor of two difference in $l_{\phi}$ obtained from magneto-conductance
and magneto-noise can arise because Eq.\ \ref{eq:ucf 2} is valid
in the case of $L_{\phi}<<L,$$L_{T}$\ \cite{adroguer2012diffusion,akkermans2007mesoscopic},
where $L_{T}$ is the thermal length. Moreover, the phase breaking
time $\tau_{\phi}$ and hence, the phase breaking length $l_{\phi}=\sqrt{D\tau_{\phi}}$\ ($D$
is the electron diffusivity) relevant for weak localization\ (WL)
is related to the Nyquist dephasing rate\ \cite{altshuler1982effects},
whereas in case of UCF, the relevant scattering time for $2$D is
the out-scattering time maybe related to the inverse of the inelastic
collision frequency and differ by a logarithmic factor compared to
the WL rate\ \cite{blanter1996electron,hoadley1999experimental,trionfi2004electronic,mcconville1993weak,shamim2017dephasing,chandrasekhar1991weak}.

In conclusion, we have measured the Fermi energy dependent aperiodic
and reproducible fluctuations in mesoscopic topological insulator
systems. The magnetic field dependence of these fluctuations conclusively
indicate that they arise from universal conductance fluctuations.
Most importantly, a factor of two reduction in noise magnitude close
to the Dirac point is observed, which provides an unambiguous proof
that the time reversal symmetry in disordered topological insulators
is intrinsically maintained.

We thank Saquib Shamim for useful discussions. We acknowledge the
Department of Science and Technology (DST) for funding.

\bibliographystyle{apsrev4-1}
%

\end{document}